\documentclass{llncs}
\usepackage{booktabs} 
\usepackage{multirow}
\usepackage{braket}
\usepackage{graphicx}
\usepackage{amssymb}
\usepackage{amsmath}
\usepackage{url}
\usepackage{rotating}
\usepackage{color}
\usepackage[table,usenames,dvipsnames]{xcolor}

\begin{document}

\title{Quality Attributes on Quantum Computing Platforms}
\titlerunning{Quality Attributes on QCPs}  
%

\author{Balwinder Sodhi\inst{1}}
\authorrunning{Balwinder Sodhi} 
%
%
\institute{Indian Institute of Technology Ropar, Nangal Rd. Ropar PB 140001 India\\
\email{sodhi@iitrpr.ac.in}, \\
WWW home page: \texttt{http://www.iitrpr.ac.in/sodhi}
}

\maketitle              

\begin{abstract}
As the practical Quantum Computing Platforms (QCPs) rapidly become a reality, it is desirable to harness their true potential in software applications. Thus it becomes important to determine the implications of QCPs for software architecture. 

In this paper we present the in-depth examination of state-of-the-art QCPs for identifying all such characteristics of a QCP that are relevant from software architecture perspective. Lack of a native quantum operating system, a hard dependency on quantum algorithms, the lower level of programming abstractions are few, out of many, examples of QCP characteristics which may affect architecture of quantum software applications. 

Key contributions of this paper include identifying: \textbf{i)} The general architecture of a QCP, \textbf{ii)} The programming model which is typically used when developing software for a QCP, \textbf{iii)} Architecturally significant characteristics of QCPs and \textbf{iv)} The impact of these characteristics on various Quality Attributes (QAs).


We show that except performance and scalability, most of the other QAs (e.g. maintainability, testability, reliability etc.) are adversely affected by different characteristics of a QCP.

\end{abstract}

\section{Introduction}
\label{sec:intro}
The idea of Quantum computers was proposed in the 1980s by Richard Feynman and Yuri Manin. However, the origin of  quantum information processing can traced back to the early 1970's work on quantum communication channels by Holevo\cite{holevo1973bounds}, and on quantum information theory by Ingarden\cite{ingarden1976quantum}. It was not until 1990s that usable algorithms which exploited quantum computing properties appeared: first due to Peter Shor for quickly factoring large integers\cite{shor1994algorithms}, and another due to Lov Grover for fast database search\cite{grover1996fast}. It was only in 2011 that the first commercially viable quantum computer was reported\cite{dwave-opens-box-2011}. Few recent developments in QCP space include Intel's delivery\cite{intel-qpu-2017} of a 17-qubit superconducting quantum chip, and that of IBMs\cite{ibm50qbit2017} 50-qubit quantum computer. It is possible today for a programmer to use a real quantum computer through cloud based quantum programming platforms (e.g. IBM Q Experience\cite{ibm-qx}).

From the perspective of software development, \textit{quantum computing} is one of the most recent paradigm shifts which is underway. In this paper we use the term \textit{Quantum Computing Platform (QCP)} to refer to the entire apparatus (hardware and software) which is necessary to develop and deploy \textit{quantum software} applications. Considering the pace of development happening in the domain of QCPs it is pertinent to investigate how will the large scale professional development of quantum software be different (or similar) to building software for classical computing platforms. Some of the important questions in this context could be:
\begin{enumerate}
\item What are the key characteristics of QCPs which are efficacious for software development?
\item In what way does a QCP affect the quality attributes (and non-functional requirements) of quantum software applications?
\item What type of software architectures are suitable for quantum software? 
\item How to assess the suitability of a QCP for a given computing problem/task?
\item How do QCPs affect SDLC activities? For instance, do the classical approaches apply as-is for testing and verification of quantum software?
\item What type of software development processes and methods would be suited for building quantum software?
\end{enumerate}

In this paper we address the \underline{first two} question via in-depth examination of the QCPs available today. 

\subsection{Method of research}
\label{sec:method}
The central idea underlying our approach for finding answers for these questions is as follows: \textit{In order to build various QAs into a software application a software architect (in)directly leverages (or mitigates) the characteristics of the target computing platform for which the software application is being built.} 
Consider this example: building the \textit{performance} QA into an application often demands exploiting parallelism present in a problem/task. However, if the target computing platform does not offer multiprocessing (e.g. if it has only one single-core CPU) then the application will not be able to truly realize the QA of performance. A proficient software architect must know: \textbf{i)} \textit{which} all characteristics of a computing platform affect different QAs for an application, and \textbf{ii)} \textit{how} are the QAs affected. 

Addressing the point-\textbf{i} (i.e. \textit{which} characteristics) is relatively straight forward -- one mainly needs to examine in detail the available literature and software artifacts of a computing platform for identifying its key characteristics. However, the point-\textbf{ii} is not as straight forward to address. We take the following two-step approach for addressing point-\textbf{ii} (i.e. determining \textit{how} a QA is affected by platform characteristics):
\begin{enumerate}
\item Examine the definition and general scenario\cite[Part-II]{bass2007software} for a QA to identify the cause-and-effect relationships\cite{bachmann2002illuminating}, if any, between platform characteristics and the QA.
\item Examine\cite{deriving-tactics,bass2000quality} the design tactics prescribed\cite[Part-II]{bass2007software} for realizing a QA to identify the above relationships.
\end{enumerate}

Application of this approach is demonstrated in \S\ref{sec:effect-swp} where we determine the impact of QCP characteristics on various QAs.

The structure of our paper broadly reflects how we carried out this study: In \S\ref{sec:q-platforms} we present the quantum computing basics that are relevant from the perspective of developing quantum software, and bring out the general architecture of a QCP. We discuss the techniques which are exploited by currently known quantum algorithms, and present the quantum programming model in \S\ref{sec:q-software-dev}. Characteristics of QCPs have been determined in \S\ref{sec:q-platform-chars} after which we discuss in \S\ref{sec:effect-swp} the effect of these characteristics on QAs.

\section{Quantum computing platforms}
\label{sec:q-platforms}
Determining software architectural aspects of quantum software development requires deeper comprehension of important differences between the classical and the quantum computers. Though a detailed treatment of theoretical underpinning of quantum phenomenon and computing is available in standard quantum computing texts such as \cite{nielsen2002quantum}, we present quantum computing concepts that are necessary for our discussion.

\subsection{What makes quantum computing different}
A classical computer stores and processes information in the form of binary bits (1 and 0). At a physical level a bit is realized using suitable properties (e.g. voltage or current etc.) of some physical device. An $n$ bit memory cell in a classical computer can potentially represent $2^n$ different \emph{symbols}, but at any moment it can represent only one of these $2^n$ possibilities. Information on a quantum computer is represented/stored using quantum bits (or qubits). These qubits are physically realized via suitable physical phenomena that obey the laws of quantum mechanics. Examples of such phenomena can be: The spin of a single electron or the configuration of an individual ion. A qubit can be \textit{observed} in one of the two \textit{basis states} which are labeled as $\ket{1}$ or a $\ket{0}$. Normally, the qubit state $\ket{1}$ corresponds to classical bit 1, and $\ket{0}$ to classical 0 bit. One of the remarkable properties of qubits, which makes quantum computers much faster than classical computers, is the number of possible states in which the qubit can \textit{exist} (different from its observable state, which can be only one of the basis states). At any instant one qubit can be in a $\ket{1}$, a $\ket{0}$ state, or any \textit{quantum superposition}\cite{nielsen2002quantum} of the two. In other words, the basis states $\ket{1}$ and $\ket{0}$ and their linear combinations $x \ket{1} + y \ket{0}$ describe the possible states of a single qubit. In contrast, one classical bit at one time can be only in one of the two possible states: a 0 or 1.

Another quantum property that is exploited by quantum computing is \textit{entanglement} of qubits. Two or more individually independent quantum objects are said to be entangled when: a) their individual behavior is random, but at the same time b) it is too strongly correlated despite each object being independent from the other. A multi-qubit state that cannot be expressed as a list of the individual constituent qubits is an entangled state. Consider the Bell State\cite{bellstate} $q_1q_2 := (\ket{00} + \ket{11})/\sqrt{2}$. It is an example of an entangled two-qubit state. There is no way of expressing it as a list of one-qubit states. Suppose you measure (along some axis) one of the qubits, say $q_1$, that comprise this entangled state. It will behave randomly: in this case $q_1$ can be $\ket{0}$ or $\ket{1}$ with equal probability. Suppose you measured $q_1$ to be $\ket{0}$ then value of $q_2$ will certainly be $\ket{0}$. That is, you can predict exactly how the other qubit, $q_2$, would behave if measured along the same axis. No unentangled state exhibits this type of perfect correlation with perfect individual randomness.

The above two properties -- \textit{quantum superposition} and \textit{quantum entanglement} -- are thus very useful resources in quantum computing.

\subsection{General architecture}
\label{sec:a-platform-arch}
\begin{figure}[thpb]
      \centering
      \includegraphics[width=\linewidth]{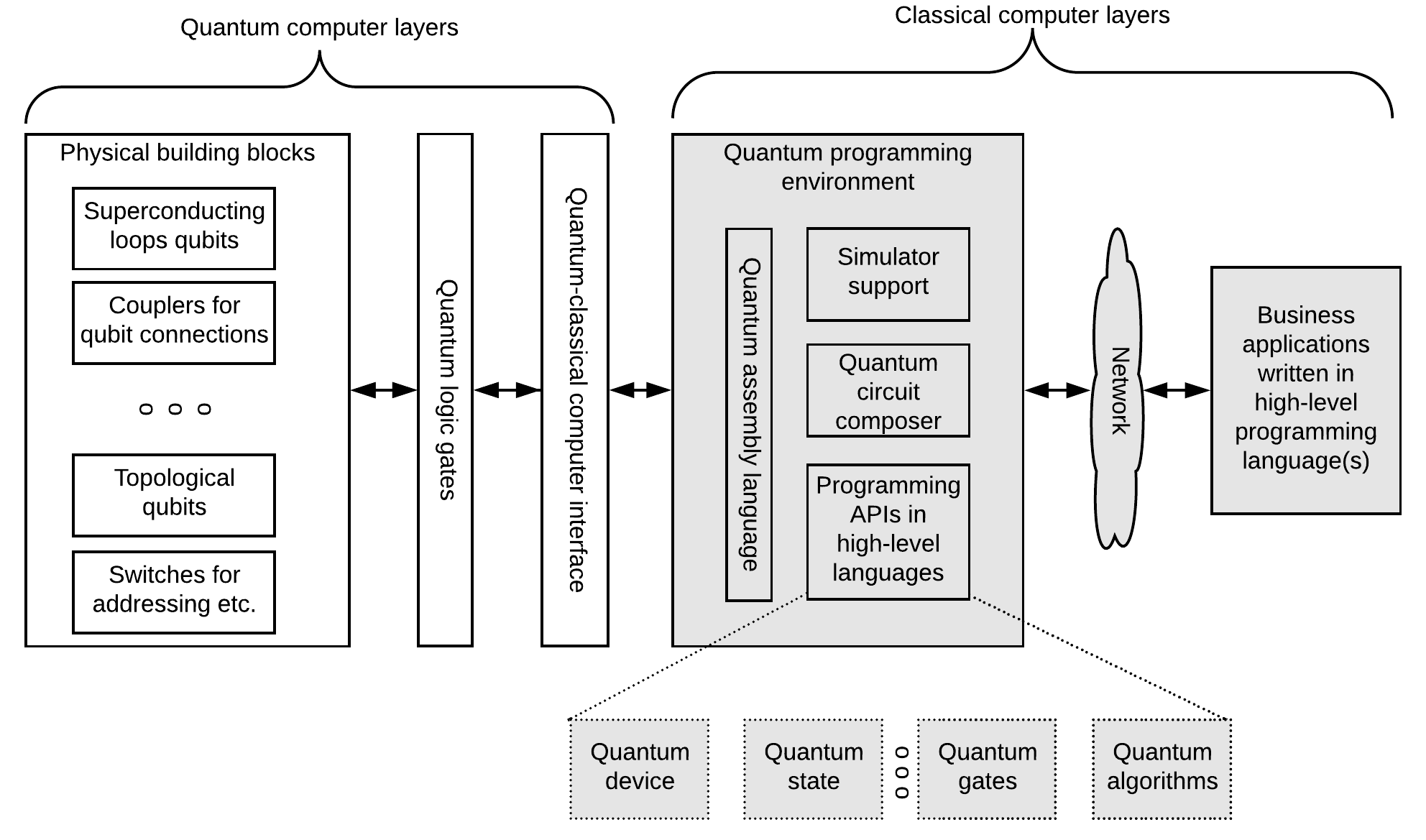}
      \caption{Architecture of quantum computing platform.}
      \label{fig:arch}
   \end{figure}

Physical quantum computing machinery of today (reminiscent of early classical computers of 1940s) are big in size\cite{dwave-hardware}. It requires special physical environment and conditions for operating correctly. The general architecture of a QCP can be depicted as shown in Fig. \ref{fig:arch}. It comprises of five layers three of which contain purely quantum hardware and circuitry and two consist of classical hardware and software:
\begin{enumerate}
  \item \textbf{Quantum layers}\cite{dwave-hardware,quantum-von-nueman-arch,ms-quantum}
  One can think of these layers as comprising the Quantum Processing Unit (QPU).
    \begin{enumerate}
	    \item \textit{Physical building blocks -- } Includes quantum hardware which typically makes use of superconducting loops for physical realization of qubits. In addition, it also contains the physical qubit coupler/interconnect circuitry among other elements which are needed for qubit addressing and control operations.
        \item \textit{Quantum logic gates -- } Physical circuitry\cite[\S 5.5]{quantum-von-nueman-arch} that makes up quantum logic gates\cite{nielsen2002quantum}.
        \item \textit{Quantum-classical interface -- } It includes the hardware and software which provides interfacing between classical computers and a QPU.
    \end{enumerate}
  \item \textbf{Classical layers}\cite{rigetti-qpu,dwave-tech-overview,ibm-qx,ms-quantum-sdk}
    \begin{enumerate}
	    \item \textit{Quantum programming environment -- } It provides items such as: i) quantum assembly language necessary for instructing a QPU, ii) the programming abstractions needed for writing quantum programs in a high-level programming language and iii) simulator support as well as IDEs etc.
        \item \textit{Business applications -- } Quantum software applications written to cater for business requirements.
    \end{enumerate}
\end{enumerate}

\section{Developing quantum software applications}
\label{sec:q-software-dev}
In preceding section we introduced the two fundamental properties -- \textit{quantum superposition} and \textit{quantum entanglement} -- which make a quantum computer much faster than classical computers at solving certain types of problems. Nature of these two properties inherently make quantum computation to be mostly probabilistic in nature. Thus, implying that quantum programs are likely to be probabilistic or randomized in nature. As such, expressing the logic or algorithm which is to be executed on a quantum computer requires special techniques and programming model.

\subsection{Quantum algorithms -- The key to harnessing power of QCPs}
\label{sec:q-problem-modeling}
Usefulness of quantum programs lies in their ability to exploit the fundamental characteristics (superposition and entanglement of qubits) of a quantum computer. Over the past few decades quantum computing researchers have developed a handful of techniques\cite{cleve1998quantum} that exploit the characteristics of quantum computers for quickly solving certain problems that take much longer to solve on a classical computer. Some of these techniques and example algorithms that exploit them are shown in Table-\ref{tab:q-algo-techniques}.

\begin{table}[h]
\caption{Techniques used by quantum algorithms}
\label{tab:q-algo-techniques}
\begin{center}
\begin{tabular}{p{5cm} p{7cm}}
\toprule
\textbf{Technique} & \textbf{Examples of target problem(s)} \\ \midrule
Amplitude amplification\cite{quantum-amplitude-amplifiy} & Quantum counting/search (Grover's algorithm\cite{grover1996fast}). \\ \hline
Quantum Fourier transform\cite{quantum-fft} & Discrete logarithm problem and the integer factorization problem in polynomial time (Shor's algorithm), and so on. \\ \hline
Phase kick-back\cite{cleve1998quantum} & Estimating Gauss sums\cite{van2002efficient} \\ \hline
Quantum phase estimation\cite{cleve1998quantum} & Estimates the phase that a unitary transformation adds to one of its eigenvectors. \\ \hline
Quantum walks\cite{quantum-walks-review,quantum-walks-algo-apps} & Element distinctness problem, Triangle-finding problem, Formula evaluation, Group commutativity \\ \hline
\end{tabular}
\end{center}
\end{table}
   
There are several algorithms available today which exploit these techniques to solve variety of computing problems much faster than a classical computer. In order to derive benefit from the capabilities of quantum computer one must map the problem at hand to one of the problems for which a quantum algorithm is known.

   \begin{figure*}[thpb]
      \centering
      \includegraphics[width=\linewidth]{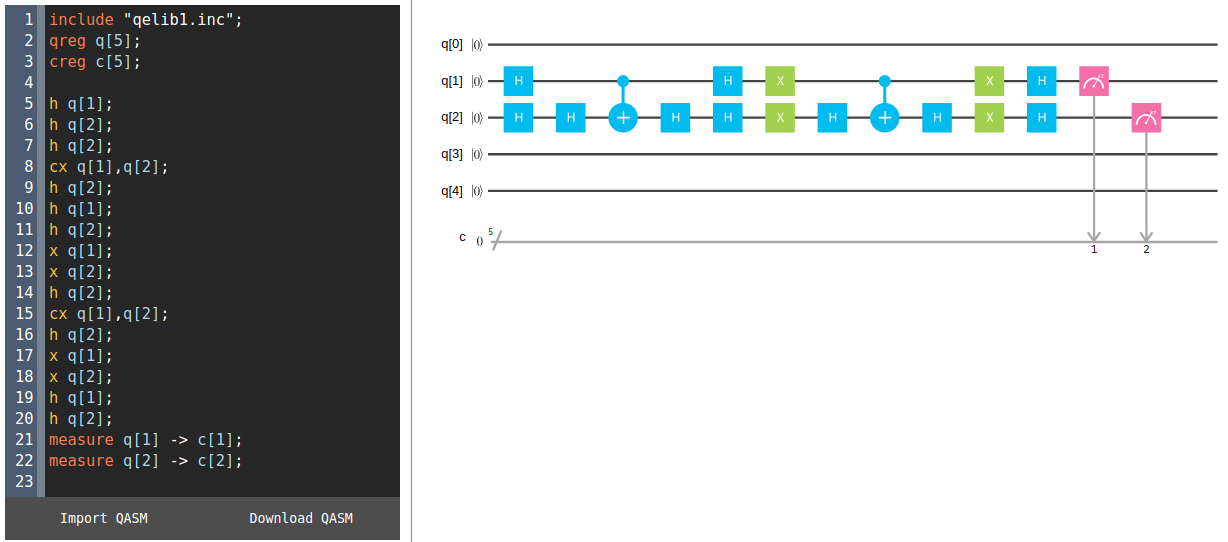}
      \caption{An example quantum circuit and corresponding Python code.}
      \label{fig:composer-grover-algo}
   \end{figure*}

Quantum circuits\cite{quantum-circuits} is one of the common models for representing quantum computation. Similar to digital logic gates which are employed by classical computers, \textit{quantum gate}s are used to compose a quantum circuit. In this model the steps of a quantum algorithm can be expressed as a sequence of quantum logic gates. Each quantum logic gate transforms the input qubits in a well defined manner, typically expressed a operations on matrices and vectors. IBM Q-Experience\cite{ibm-qx} takes this approach to expressing quantum computations. Fig. \ref{fig:composer-grover-algo} shows the screen shot of a quantum circuit (and its corresponding Python code) for implementing part of Grover's\cite{grover1996fast} algorithm on IBM Q-Experience\cite{ibm-qx} 

   \begin{figure}[thpb]
      \centering
      \includegraphics[width=0.7\linewidth]{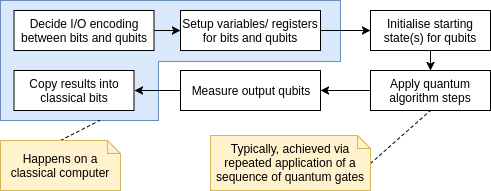}
      \caption{A quantum programming model.}
      \label{fig:q-prog-model}
   \end{figure}

\subsection{Programming model in quantum paradigm}
\label{sec:q-programming-model}
We evaluated the programming abstractions, the high-level programming language libraries that expose these abstractions and the typical steps involved in composing quantum programs on different QCPs such as \cite{ibm-qx,dwave-tech-overview,ms-quantum-sdk,rigetti-qvm} etc.

We observe that a quantum program typically comprises of parts some of which execute on a classical CPU and some on a Quantum Processing Unit (QPU). Creating such quantum programs mainly involved the following tasks:
\begin{enumerate}
\item Mapping input/output from classical bits representation to qubits.
\item Initializing the qubits state.
\item Compose the quantum circuit using suitable quantum logic gates to express steps of a quantum algorithm. The steps are repeated suitable number of times to get a reliable measurement of the outcomes.
\item Measure the output qubits state (measuring a qubit forces it to collapse into a classical bit) and transfer it to the classical bits.
\end{enumerate}

A suitable high-level programming API or instructions\cite{ms-quantum-sdk,open-qasm,rigetti-quantum-isa-2016} are typically used for composing the quantum programs. Based on these observation a quantum programming model as shown in Fig. \ref{fig:q-prog-model} has been derived. We also checked the adherence to this programming model by manually examining the sample quantum programs that are supplied by different QCPs we considered.
 
\section{Quantum software architecture}
\label{sec:q-se}

\subsection{Characteristics of a quantum computing platform}
\label{sec:q-platform-chars}
In order to understand the effect of QCPs on software architecture it is necessary to identify all those characteristics of a QCP which can affect the development, deployment, operation and management of quantum software.

Some such characteristics of the QCPs of today are:
\begin{enumerate}
\item \textbf{Platform heterogeneity --} The published technical details of different QCPs\cite{dwave-tech-overview,ibm-qx,ms-quantum,rigetti-qvm} that are available today  show the heterogeneous nature of the QCPs. Considering certain limitations such as various no-go theorems\cite{no-cloning-wootters1982,no-deleting-pati2000} that quantum phenomenon obey, dependency on classical hardware/software is inevitable. Thus, the QPUs are expected to be inherently \textit{heterogeneous} in nature. That is, both classical as well as quantum hardware and software are involved.

\item \textbf{Physical environment --} The quantum hardware requires a completely different type of physical environment. Most of the existing implementations\cite{dwave-hardware,dwave-tech-overview,ibm-qx,ms-quantum} of quantum hardware circuitry make use of superconducting loops requiring ultra-low temperatures. Achieving and maintaining such physical conditions necessitates very specialized environment for reliable operation of a QCP.

\item \textbf{Large form factor --} Due to requirement of special physical environment, quantum computers of today are very large in size. For instance, the main box of D-Wave\texttrademark 2000Q system measures\cite{dwave-hardware} roughly 10'x10'x10'.

\item \textbf{Energy efficiency -- } Energy consumption can be looked at from two main aspects: energy consumed by the QPU itself, and secondly the energy requirements for cooling and other ancillary circuitry which make up a quantum computer.
It has been observed that a quantum computer spends\cite{quantum-energy-consumption} most of its energy on cooling. A QPU by itself, however, consumes much less energy\cite{energy-efficient-qcp-2017}. As the computation speeds grow to exascale (i.e. 1000 petaflops), the energy consumption of a QPU is not expected to increase as fast as that of a CPU/GPU. Experiments conducted with D-Wave's 2000-qubit system showed\cite{dwave-energy} overall energy efficiency improvements of the order of 100x in comparison to state-of-the-art classical computing servers when considering pure computation time for running specialized algorithms.

\item \textbf{Lower level of the programming abstractions --} The programming abstractions offered by QCPs of today\cite{ibm-qx,dwave-tech-overview,ms-quantum-sdk,open-qasm} are of a low level. That is, a programmer has to work at the level of quantum logic gates (usually via a high-level language representation of it)  when expressing computational steps that he/she wants to execute via a QPU.

\item \textbf{Remote software development and deployment --} Programming tools such as IDEs, debuggers, simulators etc. that a software developer can use to create quantum software are invariably cloud based\cite{ibm-qx,dwave-tech-overview}. Only a very limited portion\cite{ms-quantum-sdk,open-qasm} of the quantum programming tools stack can be deployed and used on a programmer's local machine. For development and testing of production-ready quantum software a programmer typically requires access to a remote QCP environment.

\item \textbf{Dependency on quantum algorithms --} To exploit the real potential of quantum computers a programmer has to express the logic of his/her software using quantum algorithms. A computing task where one is looking to gain speedup by running it on a QCP is typically mapped to or broken into another task(s) for which a quantum algorithm(s) is known.

\item \textbf{Limited portability of software --} Quantum computing platforms are themselves in their infancy, though under rapid growth. As such there is lack of standards that are necessary for developing quantum programs which can be executed transparently on different QPUs. Each of the major providers of QCPs offer their proprietary programming APIs and tools\cite{dwave-tech-overview,ms-quantum-sdk,rigetti-quantum-isa-2016,rigetti-qvm,ibm-qx,open-qasm}.

\item \textbf{Limited quantum networking --} Though long distance distribution of quantum entanglement is feasible\cite{satellite-based-entanglement-distrib-2017}, realizing practical quantum communication networks is still a work in progress\cite{towards-quantum-network-simon_2017,quantum-network-kimble-2008}. As such practical quantum software which depends on availability of a reliable quantum network would be hard to achieve.

\item \textbf{Lack of native quantum operating system --} The quantum processors are still controlled via classical computing operating systems\cite{quantum-os}. We do not yet have mature multitasking and multiprocessing capabilities available for quantum processors. Most existing QCPs\cite{ibm-qx,dwave-tech-overview,ms-quantum-sdk,ms-quantum} expose the quantum gates and qubits for direct manipulation by programmers. Mature protocols and APIs that implement, for example, practical timesharing of a Quantum Processing Unit (QPU) are not available yet\cite{quantum-os}. Similarly, quantum algorithms which may exploit multiple QPUs in parallel is yet to be explored.  

\item \textbf{Limited multitasking and multiprocessing --} This follows from the lack of native quantum operating systems. A programmer has to rely on classical computer's OS for achieving any type of multitasking and multiprocessing on a given set of QPUs.

\item \textbf{Fundamentally different programming model --} As discussed in \S\ref{sec:q-software-dev}, quantum programs are inherently probabilistic. A programmer looking to harness power of a QCP must identify or design suitable quantum algorithms which are can solve the problem at hand.

\item \textbf{Dependency on classical storage --} Though limited time storage of entangled qubits is feasible\cite{quantum-storage-2011,quantum-storage-2014}, long term persistence of qubits in passive media is still not possible. Besides, it is not feasible to store arbitrary non-entangled qubits due to different no-go theorems such as no-teleportation\cite[Page 128]{pathak2013elements}, no-cloning\cite{no-cloning-wootters1982} and no-deleting\cite{no-deleting-pati2000} theorems. Thus, permanent persistence of critical data in quantum programs still requires classical storage devices.
\end{enumerate}

\subsection{Effect on quality attributes}
\label{sec:effect-swp}
Quality attributes (QA) may be described as the factors which have system-wide impact on an application's architecture, implementation as well as operation\cite{bass2007software}. The QAs that are of concern for a majority of applications may be categorized depending on whether they affect design-, runtime-, system- or user- qualities of the application\cite{swebokv3,ms-arch-guide}. 

The overall design and quality of a large majority of software applications can be considered good when the applications possess a reasonable level of at least the following QAs\cite{bass2007software,bass2000quality,qa-models-ISO-IEC}: Performance, Reliability, Scalability, Security and Usability. However, based on our experience, we considered a slightly expanded list of QAs when examining how they are affected on a QCP:

\begin{multicols}{3}
\begin{enumerate}
    \item Availability
    \item Interoperability
    \item Maintainability
    \item Manageability
    \item Performance
    \item Reliability
    \item Scalability
    \item Security 
    \item Testability
    \item Usability
\end{enumerate}
\end{multicols}

In subsequent paragraphs we discuss only those  aspects of a QA that are relevant for identifying how the QA is affected by various characteristics of QCPs\footnote{A detailed description and properties of individual QAs is readily available in standard software architecture literature such as \cite{bass2007software,ms-arch-guide,qa-models-ISO-IEC} etc.}. We take each of the QCP characteristics identified in \S\ref{sec:q-platform-chars} in turn and discuss how it affects various QAs under consideration. A characteristic may be considered \textit{favorable, unfavorable} or \textit{neutral} for building a QA into an application. Summary of the QA impact of various characteristics of QCPs is shown in Table-\ref{tab:qa-impact}.

\begin{table}[!b]
\caption{Impact$^*$ of QCP characteristics on QAs}
\label{tab:qa-impact}
\begin{center}
\begin{tabular}{r *{10}{|p{1.5 em}}}


\textbf{Characteristics~~} & 
\rotatebox{90}{Availability} & 
\rotatebox{90}{Interoperability} &
\rotatebox{90}{Maintainability} &
\rotatebox{90}{Manageability} &
\rotatebox{90}{Performance} &
\rotatebox{90}{Reliability} &
\rotatebox{90}{Scalability} &
\rotatebox{90}{Security} &
\rotatebox{90}{Testability} &
\rotatebox{90}{Usability} \\

\toprule
Platform heterogeneity~~& 
U & -- & U & U & -- & U & -- & -- & U & -- \\
Special physical environment~~ & 
U & -- & -- & U & -- & U & U & U & U & -- \\
Large form factor~~ & 
-- & -- & -- & -- & -- & -- & U & -- & -- & -- \\
Higher energy efficiency~~ & 
-- & -- & -- & -- & F & -- & F & -- & -- & -- \\
Lower level of the programming abstractions~~ &
U & -- & U & -- & -- & U & -- & -- & U & -- \\
Remote software development and deployment~~ &
-- & -- & U & -- & -- & -- & -- & -- & U & -- \\
Dependency on quantum algorithms~~ & 
-- & U & U & -- & F & -- & F & -- & U & -- \\
Limited portability of software~~ &
U & U & U & -- & -- & -- & U & -- & -- & -- \\
Limited quantum networking~~ & 
U & -- & -- & -- & U & U & U & -- & -- & -- \\
Lack of native quantum operating system~~ &
-- & -- & -- & U & U & U & U & U & -- & -- \\
Fundamentally different programming model~~ &
-- & U & U & -- & -- & U & -- & U & U & -- \\
Dependency on classical storage~~ &
-- & -- & -- & U & U & U & U & -- & -- & -- \\
\bottomrule
\multicolumn{11}{l}{$^*$Cell value indicates impact on QA: (F)avorable, (U)nfavorable, ``--'' $\rightarrow$ Unknown/Neutral}
\end{tabular}
\end{center}
\end{table}

\subsubsection{Discussion}
\label{sec:qa-impact-discussion}

\begin{enumerate}
\item \textbf{Platform heterogeneity:}
Heterogeneity makes it difficult to implement high cohesion in software. Therefore, QAs such as maintainability, reliability, robustness, reusability, and understandability get adversely affected due to low cohesion. Heterogeneous environment also means more number of disparate elements (software and hardware) to managed thus adversely affecting manageability and testability of the system.

\item \textbf{Special physical environment:}
A highly specialized physical environment needed by a QCP is difficult to create, maintain and operate. Effect of ambient noise/interference is more pronounced in case of QPUs than it is on CPUs. Such properties have adversely affect on the QAs such as availability (e.g. due to ``brittle'' nature of physical qubits), manageability, scalability (e.g. due to difficulties in quickly launching additional QPU instances) and testability of the system.

\item \textbf{Large form factor:} Most of the QAs, except scalability, remain unaffected by this property of QCPs. Large form factor makes it difficult to augment the capabilities of a QCP thus adversely affecting scalability.

\item \textbf{Higher energy efficiency:}
Except improving performance and scalability, this property does not have significant impact on most other QAs.

\item \textbf{Lower level of the programming abstractions:}
Programming the QCPs requires one to work at the level of quantum gates\cite{dwave-tech-overview,ibm-qx,ms-quantum-sdk} to compose the necessary quantum circuits for implementing steps of a quantum algorithm. Working at such low levels of abstractions is not easy and increases complexity of the code. Moreover, there are not many (as of today at least) expert quantum programmers.

This in turn adversely affects QAs such as maintainability, testability, reliability and availability.

\item \textbf{Remote software development and deployment:}
A major implication of QCPs on testing and debugging arises due to the non-local nature of a real quantum computer. The development tools/environment used for a QCP are sort of distributed. The decentralized (typically, offered via a cloud based IDE) and heterogeneous nature of the development environment will make programing, testing and debugging of quantum programs slower and tedious\cite{cloud-based-development}. This adversely affects maintainability and testability.

\item \textbf{Dependency on quantum algorithms:}
Probabilistic nature of quantum computations and the results they produce adversely affect testability and interoperability (with classical software). Synthesizing realistic pairs of $\langle input, expected~output \rangle$ for test case scenarios as well as reproducing the defects that one wants to debug requires special approaches.

Moreover, there are very few\cite{quantum_algo_progress_2004} -- only about a dozen -- quantum algorithms known for different types of problems. Software engineers have to map their problems to one of the existing few quantum algorithms. This adversely affects the ability to perform enhancement and corrective maintenance.

\item \textbf{Limited portability of software:}
Quantum computing, though rapidly evolving, is still in its infancy. It lacks standardization in several areas ranging from high-level programming APIs to low-level hardware. For example, a quantum program written using Rigetti's quantum ISA\cite{rigetti-quantum-isa-2016} may not be executable on Open QASM\cite{open-qasm} supported by IBM. As such the portability of software is adversely affected on QCPs. Lack of portability in turn adversely affects availability, interoperability, maintainability and scalability.

\item \textbf{Limited quantum networking:}
A reliable quantum network is necessary for building reliable and high-performance quantum software. Quantum networks not yet production ready \cite{towards-quantum-network-simon_2017,quantum-network-kimble-2008}. Thus, the performance, scalability, reliability and, in turn, availability will be adversely affected. 

\item \textbf{Lack of native quantum operating system:} 
A native operating system helps in harnessing full potential of a hardware in a secure and effective manner. This is lacking\cite{quantum-os} in case of QCPs. It prohibits, for instance, practical timesharing of a QPU. Thus, the performance, manageability, reliability, scalability and security will be adversely affected.

\item \textbf{Fundamentally different programming model:} Quantum programs are inherently probabilistic and requires a fundamentally different approach to programming (please see \S\ref{sec:q-programming-model} ). This affects the ease of use of the underlying technology and in turn the code complexity both of which are important factors that influence\cite{sw-maint-economics-2006} development and maintenance of dependable software.

As such a QCP adversely affects maintainability, interoperability, security and testability QAs.

\item \textbf{Dependency on classical storage:} 
For durably persisting critical data the quantum programs depend on classical storage devices due to different no-go theorems such as no-teleportation\cite[Page 128]{pathak2013elements}, no-cloning\cite{no-cloning-wootters1982} and no-deleting\cite{no-deleting-pati2000} theorems. On a QCP this adversely affects manageability, performance and scalability.

\end{enumerate}

\subsection{Threats to validity}
\label{sec:threats2v}
Quantum computing is a fast evolving area of technology. The characteristics of QCPs that we have identified are based on the study of currently available state-of-the-art quantum hardware and software. We expect advances in quantum computing will invalidate at least few of these characteristics in coming years. For example, the production-ready native quantum operating systems are likely to be feasible in the years to come.

Further, the list of characteristics that we have given is not an exhaustive list. It is quite likely that there are additional QCP characteristics which, taken together, may be of significance in specific software development scenarios. Another point that we would like to highlight is that our findings are derived from: a) published information about QCPs and b) experimental programming on the QCPs accessible to us. It is likely that there are unpublished features/information about those QCPs which can affect the software architecture of quantum software applications. Next, we have not covered every QA which is relevant for a wider set of application types. There may be QAs such as auditability, distributability, extensibility etc. which are relevant and may be affected by the QCPs. Lastly, our own experience and know-how as software architects has influenced to some extent what we identified as the effect of different QCPs on various QAs.

\section{Conclusions}
Programmers' interest in applying quantum computing has surged in the recent past. Leveraging this technology in solving general purpose business problems requires deeper understanding of important characteristics of QCPs, particularly those which are relevant for software development. 

We have shown that the key characteristics of a QCP make it different from a classical computing platform. For instance, availability and know-how of quantum algorithm(s) for solving a task at hand are a hard requirement for developing quantum programs. Such characteristics of a QCP affect various Quality Attributes (QAs) of the quantum application software. The QCP characteristics such as limited portability of quantum programs, lack of native quantum operating system services, dependency on quantum algorithms and so on, adversely affect the ability to realize various QAs such as scalability, portability, testability and maintainability etc. of the quantum application software. The QAs that are favorably impacted by QCP characteristics include performance and scalability.

Overall, the specialized nature of QCPs appears to have a significant implication: it limits the use of QCPs for very specialized application areas where \textit{performance} QA is of chief importance. However, because the quantum computing is undergoing rapid development we expect that the evolution of this technology will likely introduce additional concerns and factors which may affect the architecture of quantum software applications.

%
%
\bibliographystyle{unsrt}
\bibliography{quantum}

\end{document}